\documentclass[%
prd,
 reprint,
superscriptaddress,
nofootinbib,
 amsmath,amssymb,
 siunitx,
 aps,
]{revtex4-2}

\usepackage{graphicx}
\usepackage{dcolumn}
\usepackage{bm}
\usepackage{xcolor}

\usepackage{hyperref}
\hypersetup{
    colorlinks,
    linkcolor={blue!50!black},
    citecolor={blue!50!black},
    urlcolor={blue!50!black}
}

\usepackage[left]{lineno}

\begin{document}

\title{Ionization yield in a methane-filled spherical proportional counter}

\author{M.~M.~Arora}
\affiliation{Department of Mechanical and Materials Engineering, Queen’s University, Kingston, Ontario K7L 3N6, Canada}
\author{L.~Balogh}
\affiliation{Department of Mechanical and Materials Engineering, Queen’s University, Kingston, Ontario K7L 3N6, Canada}
\author{C.~Beaufort}
\affiliation{LPSC-LSM, Universit\'{e} Grenoble-Alpes, CNRS-IN2P3, Grenoble, 38026, France}
\author{A.~Brossard}
\altaffiliation[now at ]{TRIUMF,  Vancouver,  BC  V6T  2A3,  Canada}
\affiliation{Department of Physics, Engineering Physics and Astronomy, Queen’s University, Kingston, Ontario, K7L 3N6, Canada}
\author{M.~Chapellier}
\affiliation{Department of Physics, Engineering Physics and Astronomy, Queen’s University, Kingston, Ontario, K7L 3N6, Canada}
\author{J.~Clarke}
\affiliation{Department of Physics, Engineering Physics and Astronomy, Queen’s University, Kingston, Ontario, K7L 3N6, Canada}
\author{E.~C.~Corcoran}
\affiliation{Chemistry and Chemical Engineering Department, Royal Military College of Canada, Kingston, Ontario K7K 7B4, Canada}
\author{J.-M.~Coquillat}
\affiliation{Department of Physics, Engineering Physics and Astronomy, Queen’s University, Kingston, Ontario, K7L 3N6, Canada}
\author{A.~Dastgheibi-Fard}
\affiliation{LPSC-LSM, Universit\'{e} Grenoble-Alpes, CNRS-IN2P3, Grenoble, 38026, France}
\author{Y.~Deng}
\affiliation{Department of Physics, University of Alberta, Edmonton, Alberta T6G 2E1, Canada}
\author{D.~Durnford}
 \email[e-mail:]{ddurnfor@ualberta.ca}
\affiliation{Department of Physics, University of Alberta, Edmonton, Alberta T6G 2E1, Canada}
\author{C.~Garrah}
\affiliation{Department of Physics, University of Alberta, Edmonton, Alberta T6G 2E1, Canada}
\author{G.~Gerbier}
\affiliation{Department of Physics, Engineering Physics and Astronomy, Queen’s University, Kingston, Ontario, K7L 3N6, Canada}
\author{I.~Giomataris}
\affiliation{IRFU, CEA, Universit\'{e} Paris-Saclay, F-91191 Gif-sur-Yvette, France}
\author{G.~Giroux}
\affiliation{Department of Physics, Engineering Physics and Astronomy, Queen’s University, Kingston, Ontario, K7L 3N6, Canada}
\author{P.~Gorel}
\affiliation{SNOLAB, Lively, Ontario, P3Y 1N2, Canada}
\author{M.~Gros}
\affiliation{IRFU, CEA, Universit\'{e} Paris-Saclay, F-91191 Gif-sur-Yvette, France}
\author{P.~Gros}
\affiliation{Department of Physics, Engineering Physics and Astronomy, Queen’s University, Kingston, Ontario, K7L 3N6, Canada}
\author{O.~Guillaudin}
\affiliation{LPSC, Universit\'{e} Grenoble-Alpes, CNRS-IN2P3, Grenoble, 38026, France}
\author{E.~W. Hoppe}
\affiliation{Pacific Northwest National Laboratory, Richland, Washington 99354, USA}
\author{I.~Katsioulas}
\altaffiliation[now at ]{European Spallation Source ESS ERIC (ESS), Lund, SE-221 00, Sweden}
\affiliation{School of Physics and Astronomy, University of Birmingham, Birmingham, B15 2TT, United Kingdom}
\author{F.~Kelly}
\affiliation{Chemistry and Chemical Engineering Department, Royal Military College of Canada, Kingston, Ontario K7K 7B4, Canada}
\author{P.~Knights}
\affiliation{School of Physics and Astronomy, University of Birmingham, Birmingham, B15 2TT, United Kingdom}
\author{P.~Lautridou}
\affiliation{SUBATECH, IMT-Atlantique/CNRS-IN2P3/Nantes University, Nantes, 44307, France}
\author{A.~Makowski}
\affiliation{Department of Physics, Engineering Physics and Astronomy, Queen’s University, Kingston, Ontario, K7L 3N6, Canada}
\author{I.~Manthos}
\affiliation{School of Physics and Astronomy, University of Birmingham, Birmingham, B15 2TT, United Kingdom}
\affiliation{Institute for Experimental Physics, University of Hamburg, Hamburg, 22767, Germany}
\author{R.~D.~Martin}
\affiliation{Department of Physics, Engineering Physics and Astronomy, Queen’s University, Kingston, Ontario, K7L 3N6, Canada}
\author{J.~Matthews}
\affiliation{School of Physics and Astronomy, University of Birmingham, Birmingham, B15 2TT, United Kingdom}
\author{H.~M.~McCallum}
\affiliation{Chemistry and Chemical Engineering Department, Royal Military College of Canada, Kingston, Ontario K7K 7B4, Canada}
\author{H.~Meadows}
\affiliation{Department of Physics, Engineering Physics and Astronomy, Queen’s University, Kingston, Ontario, K7L 3N6, Canada}
\author{L.~Millins}
\affiliation{School of Physics and Astronomy, University of Birmingham, Birmingham, B15 2TT, United Kingdom}
\affiliation{Particle Physics Department, STFC Rutherford Appleton Laboratory, Chilton, Didcot, OX11 OQX, United Kingdom}
\author{J.-F.~Muraz}
\affiliation{LPSC-LSM, Universit\'{e} Grenoble-Alpes, CNRS-IN2P3, Grenoble, 38026, France}
\author{T.~Neep}
\affiliation{School of Physics and Astronomy, University of Birmingham, Birmingham, B15 2TT, United Kingdom}
\author{K.~Nikolopoulos}
\affiliation{School of Physics and Astronomy, University of Birmingham, Birmingham, B15 2TT, United Kingdom}
\affiliation{Institute for Experimental Physics, University of Hamburg, Hamburg, 22767, Germany}
\author{N.~Panchal}
\affiliation{Department of Physics, Engineering Physics and Astronomy, Queen’s University, Kingston, Ontario, K7L 3N6, Canada}
\author{M.-C.~Piro}
\affiliation{Department of Physics, University of Alberta, Edmonton, Alberta T6G 2E1, Canada}
\author{N.~Rowe}
\affiliation{Department of Physics, Engineering Physics and Astronomy, Queen’s University, Kingston, Ontario, K7L 3N6, Canada}
\author{D.~Santos}
\affiliation{LPSC-LSM, Universit\'{e} Grenoble-Alpes, CNRS-IN2P3, Grenoble, 38026, France}
\author{G.~Savvidis}
\affiliation{Department of Physics, Engineering Physics and Astronomy, Queen’s University, Kingston, Ontario, K7L 3N6, Canada}
\author{I.~Savvidis}
\affiliation{Aristotle University of Thessaloniki, Thessaloniki, 54124 Greece}
\author{D.~Spathara}
\affiliation{School of Physics and Astronomy, University of Birmingham, Birmingham, B15 2TT, United Kingdom}
\author{F.~Vazquez~de~Sola~Fernandez}
 \altaffiliation[now at ]{Nikhef (Nationaal instituut voor subatomaire fysica), Science Park 105, 1098 XG Amsterdam, Netherlands}
\affiliation{SUBATECH, IMT-Atlantique/CNRS-IN2P3/Nantes University, Nantes, 44307, France}
\author{R.~Ward}
\affiliation{School of Physics and Astronomy, University of Birmingham, Birmingham, B15 2TT, United Kingdom}
\affiliation{Institute for Experimental Physics, University of Hamburg, Hamburg, 22767, Germany}

\collaboration{NEWS-G Collaboration}

\date{\today}

\begin{abstract}
Spherical proportional counters (SPCs) are gaseous particle detectors sensitive to single ionization electrons in their target media, with large detector volumes and low background rates. The $\mbox{NEWS-G}$ collaboration employs this technology to search for low-mass dark matter, having previously performed searches with detectors at the Laboratoire Souterrain de Modane (LSM), including a recent campaign with a 135 cm diameter SPC filled with methane. While \emph{in situ} calibrations of the detector response were carried out at the LSM, measurements of the mean ionization yield and fluctuations of methane gas in SPCs were performed using a $30\,\mathrm{cm}$ diameter detector. The results of multiple measurements taken at different operating voltages are presented. A UV laser system was used to measure the mean gas gain of the SPC, along with $\mathrm{^{37}Ar}$ and aluminum-fluorescence calibration sources. These measurements will inform the energy response model of future operating detectors.
\end{abstract}\maketitle


\section{\label{sec:intro}Introduction}

The $\mbox{NEWS-G}$ collaboration \cite{website} searches for low-mass dark matter (DM) with spherical proportional counters (SPCs) \cite{Arnaud2018}. Particularly well-adapted to search for nuclear recoils from WIMP-like particle DM with masses below a few $\mathrm{GeV/c^2}$, gas mixtures are used as targets, typically including neon and methane. The detector has sensitivity to single-electron ionizations. 

In the fall of 2019, the $\mbox{NEWS-G}$ collaboration deployed a $135\,\mathrm{cm}$ diameter SPC at the Laboratoire Souterrain de Modane (LSM). A DM physics campaign was completed with approximately 10 days of data taken with $135\,\mathrm{mbar}$ of methane \cite{newsg_lsm,newsg_detector,newsg_copper}, employing the novel \mbox{``ACHINOS''} multianode sensor \cite{newsg_achinos,Giganon_2017}. Methane is an appealing target for DM searches, with a high fraction of low atomic mass elements and a low cross section for Compton scattering of background photons. Additionally, atomic hydrogen has the largest proton spin-dependent coupling to DM.

Careful characterization of the ionization properties of methane is required to conduct a DM search. To this end, a UV laser system was used for \emph{in situ} measurements of the gas gain characteristics of the detector. A gaseous $^{37}\mathrm{Ar}$ radioactive source was used to measure the energy response of electronic recoil events uniformly in the detector volume \cite{newsg_detector,newsg_laser, newsg_ar37}. Additionally, a source of aluminum fluorescence---induced by $^{241}\mathrm{Am}$ decays---was placed on the inner cathode surface of the detector. Both types of calibration measurements are required to estimate the ionization yield of electronic recoils in the gas, an incident particle energy-dependent quantity, $W(E)$, defined as the mean energy required to produce one electron-ion pair. This quantity is specific to different gases and depends on the species of the incident particle, with the yield of nuclear recoils being reduced compared to that of electronic recoils by the quenching factor \cite{lindhard,newsg_tunl,katsioulas}. A typical parametrization of this energy dependence is

\begin{equation}
    W(E) = E \times \frac{W_a}{E-U},
    \label{eq:w_inokuti}
\end{equation}

\noindent where $U$ is a low energy cutoff, typically close to the average energy of subionization excitation levels, and $W_a$ is the asymptotic limit of $W(E)$ for $E \gg U$ \cite{inokuti}. This model has been widely used and has shown general agreement with measurements in various gases down to sub-keV energies \cite{IAEA_1995,ICRU_rep31}.

For low-mass DM searches---for which the expected signal is predominantly single or few ionization electron events---knowledge of the ionization yield of the detector at low energy is critical. In this work, independent measurements of the ionization yield of methane in an SPC were performed to complement the \emph{in situ} measurements carried out at the LSM \cite{newsg_lsm}. Data was taken in pure methane using a $30\,\mathrm{cm}$ diameter SPC at Queen's University, employing a single anode sensor with measurements at different operating voltages. The following sections describe the experimental setup (Sec.~\ref{sec:setup}), laser calibrations of the SPC (Sec.~\ref{sec:laser}), modeling of the decay of $\mathrm{^{37}Ar}$ (Sec.~\ref{sec:ar_decay}), and ionization yield measurement results (Sec.~\ref{sec:wvalue_fits}).

\section{\label{sec:setup}Experimental setup}

\begin{figure*}
\center
\includegraphics[width=0.52\textwidth]{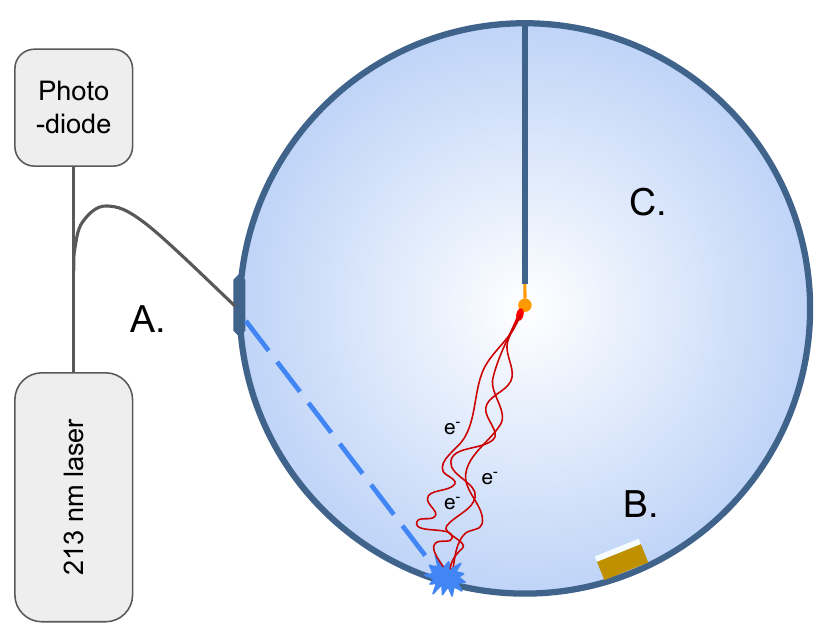}
\includegraphics[width=0.46\textwidth]{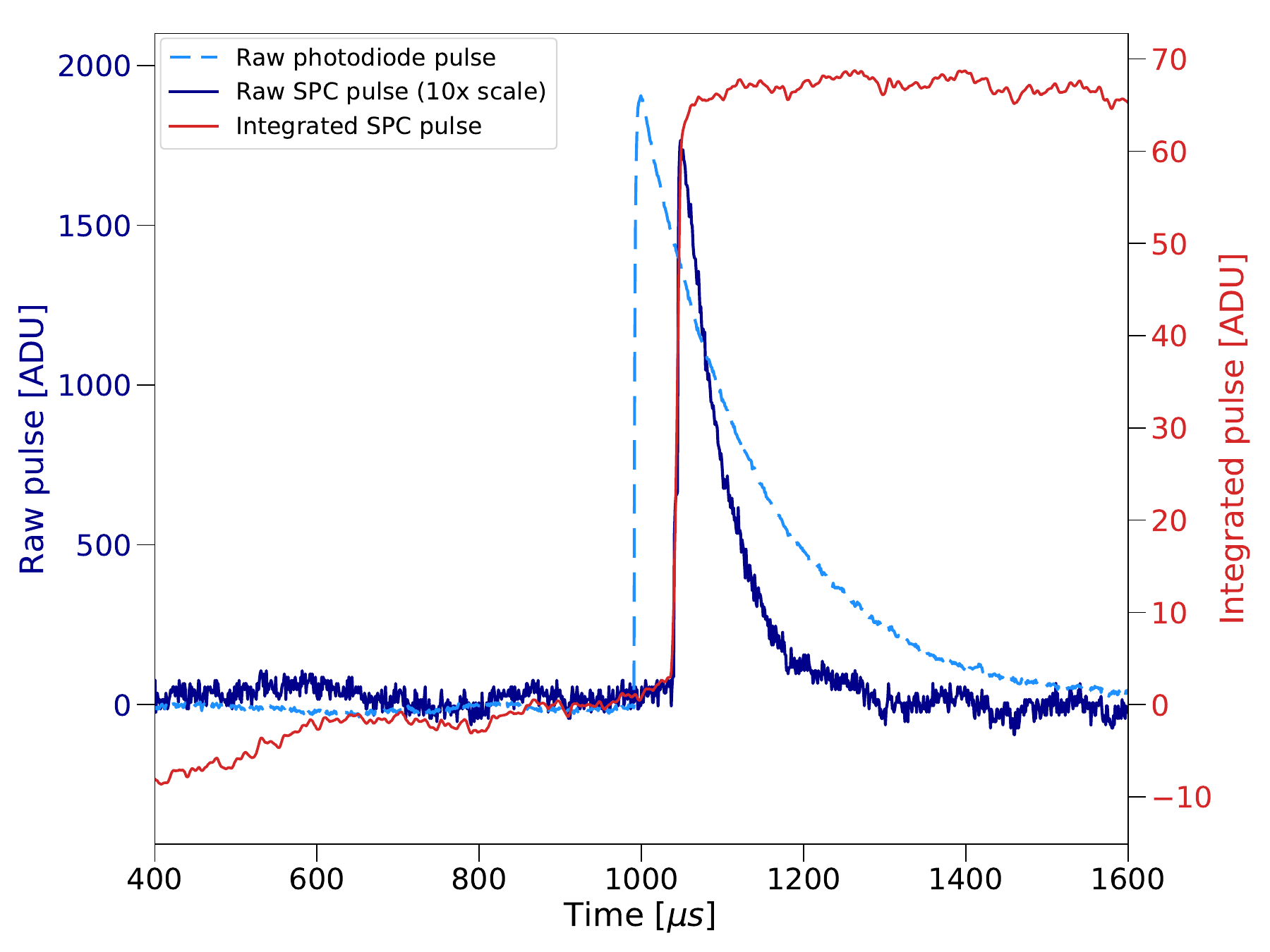}
  \caption{\label{fig:Setup} Left: Adapted from Fig.\ 1 of Ref.\ \cite{newsg_laser}. A schematic of the experimental setup including A.\ the pulsed $213\,\mathrm{nm}$ laser directed into a photodiode (PD) and the SPC where it extracts photoelectrons from the inner surface of the vessel, B.\ the aluminum-wrapped $\mathrm{^{241}Am}$ source producing fluorescence x-rays inside the detector, and C.\ the gaseous $\mathrm{^{37}Ar}$ calibration source throughout the entire detector volume. Right: An example of the raw PD signal (blue dashed curve) produced by a UV laser pulse, taken at an operating voltage of $1250\,\mathrm{V}$. The induced signal in the SPC is shown as well, including the raw pulse (dark blue curve) and the treated, integrated pulse (red curve). Both the PD and SPC raw pulses share a common y-axis, with the SPC pulse being scaled by a factor of $10$ for ease of interpretation. The SPC signal in this example likely corresponds to 1--2 photoelectrons, and is slightly delayed relative to the PD pulse by the drift time of the electrons through the SPC.}
\end{figure*}

A $30\,\mathrm{cm}$ diameter stainless steel SPC was used for these measurements. The sensor comprised a $2\,\mathrm{mm}$ diameter spherical anode and a Bakelite field-shaping component \cite{newsg_laser}. A simplified illustration of the setup is presented in Fig.\ \ref{fig:Setup}. The vessel was filled with $50\,\mathrm{mbar}$ of methane and operated at a range of stable anode voltages, from 1.2 to $1.3\,\mathrm{kV}$. The repeated measurements at different voltages serve as a cross-check, since the ionization yield should not inherently depend on the operating voltage.


The raw SPC signal is amplified by a Canberra model 2006 charge-sensitive preamplifier \cite{canberra}, then digitized at $1042\,\mathrm{kHz}$ by a custom DAQ board and software \cite{ali_cali}, with a trigger on the derivative of the signal for data acquisition. Pulse treatment primarily constitutes deconvolving the signal for the amplifier response and the effect of the motion of the avalanche ions \cite{Giomataris_2008,newsg_axion}. Within each $2\,\mathrm{ms}$ event window, the processed pulse is integrated within a central $75\,\mathrm{\mu s}$ subwindow, encompassing the signal of the ionization electrons, which the DAQ software centers in the event window. This subwindow size is defined to be larger than the average spread in ionization electron arrival time in the given detector conditions, while not being so large as to incorporate unnecessary baseline noise. This integrated pulse is then used to define the amplitude and risetime (time between 10 and 90\% of the maximum amplitude being reached) for each event. An example of a treated few-electron pulse is shown in Fig.\ \ref{fig:Setup}. There is a nonlinear, monotonically increasing relationship between risetime and the radial position of the event, as primary electrons originating far from the anode will experience more diffusion as they drift, resulting in a larger risetime \cite{Arnaud2018}.

To perform ionization yield measurements across the energy regime of interest for DM searches, two calibration sources were used. An $\mathrm{^{241}Am}$ source wrapped in aluminum was placed inside the SPC (see Fig.\ \ref{fig:Setup}). The $\alpha$ particles from $\mathrm{^{241}Am}$ \cite{nndcAm241} induce $1.486 \,\mathrm{keV}$ x-ray fluorescence from the aluminum \cite{relax}, which interact primarily in the gas via the photoelectric effect. The second source used was gaseous $\mathrm{^{37}Ar}$ injected into the detector volume. $\mathrm{^{37}Ar}$ decays via electron capture with a relatively short $35.04$ day half-life \cite{Ar37_5}, emitting x-rays and Auger electrons. The $\mathrm{^{37}Ar}$ used was produced with the SLOWPOKE-II research reactor at the Royal Military College of Canada by irradiating calcium powder to produce the radio-isotope via the reaction $\mathrm{^{40}Ca(n,\alpha)^{37}Ar}$ \cite{newsg_ar37}. 

The detector's ionization electron response was calibrated \emph{in situ} with a $213\,\mathrm{nm}$ pulsed laser setup \cite{newsg_laser}, shown in Fig.\ \ref{fig:Setup}. The UV laser light is shone into the sphere through a fiber-optic feedthrough flange. The gas is transparent at this wavelength, but electrons are extracted from the interior metal surface of the sphere. A split fiber is used to direct laser light at a Thorlabs DET10A silicon photodiode (PD) \cite{thorlabs} used to tag laser events. When collecting laser calibration data, the acquisition trigger is based on the PD signal, which has an approximately 100\% trigger efficiency independently of the pulse intensity. 

\section{\label{sec:laser}Laser Calibration}

\begin{figure}
  \centering
  \includegraphics[width=0.48\textwidth]{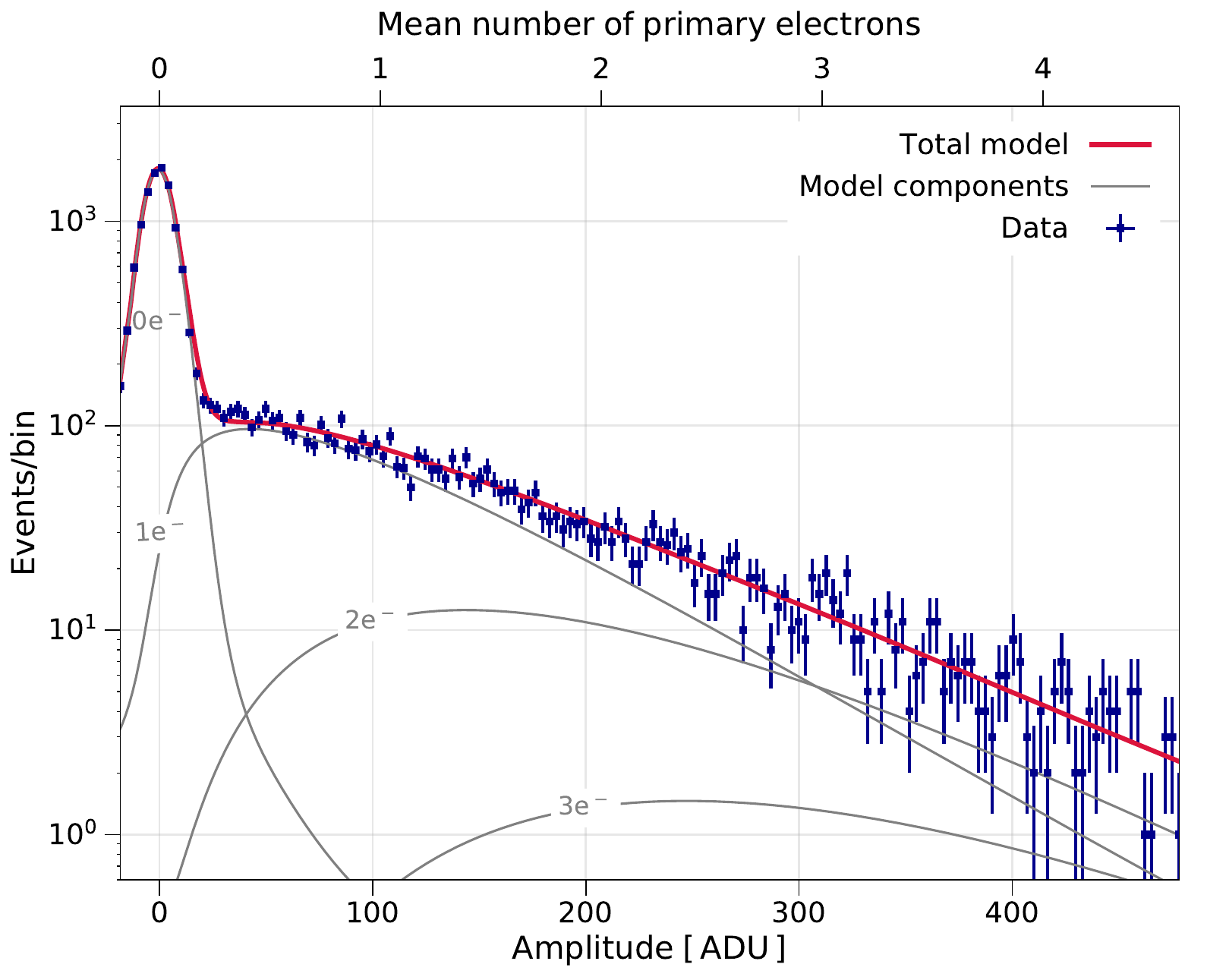}
  \caption{\label{fig:wvalue_laser} Example of a UV laser calibration result, showing the amplitude spectrum of tagged laser events (blue histogram) and resulting fit (red curve). The model has contributions from null (noise) events, and 1, 2, and 3 electron events (gray curves). The parameter estimations from this fit are $\theta = 0.64 \, \pm \, 0.06$ and $\left \langle G \right \rangle = 103.6 \,\pm \, 1.5\,\mathrm{ADU}$. The p-value of the fit is $0.29$ based on the $\chi^2$ value of the fit compared to the distribution of $\chi^2$ for MC data with the same best-fit parameters.}
\end{figure}

The detector response for a single electron depends on the exact anode and SPC geometry, the voltage applied, and gas conditions. Therefore, measurements of the mean gas gain and fluctuations are required for each experimental setup. Avalanche statistics for the number of multiplication electrons are typically modeled with the Polya distribution \cite{L11,L12,L13,L14}, with one shape parameter $\theta$ and a mean gain $\left \langle G \right \rangle$, which has units of $\mathrm{ADU / e^-}$ (analog-to-digital units per primary electron). As a function of avalanche yield $A$---which can represent the number of avalanche electrons or its size in energy-equivalent units---the probability distribution function is given as \cite{newsg_laser}

\begin{equation}
\begin{aligned}
    P_{\mathrm{Polya}}\left(A | \theta, \left \langle G \right \rangle \right) & = \frac{1}{\left \langle G \right \rangle} \frac{\left(1 + \theta \right)^{1 + \theta}}{\Gamma \left(1 + \theta \right)} \left( \frac{A}{\left \langle G \right \rangle} \right)^{\theta} \\ & \times \exp \left( -\left(1 + \theta \right)\frac{A}{\left \langle G \right \rangle} \right).
\end{aligned}
\label{polya_eq}
\end{equation}

The response for $N$ primary electrons is obtained by summing $N$ random numbers drawn from this distribution, which corresponds to the $N$th convolution of Eq.\ (\ref{polya_eq}) with itself $P_{\mathrm{Polya}}^{(N)}$, the analytical form for which is given in Ref.\ \cite{newsg_laser}. The number of primary electrons reaching the anode at a given laser intensity was modeled with a Poisson distribution with mean $\mu$, since for each UV photon incident to the surface there is a given probability of extracting an electron. Hence, the probability for obtaining $N$ primary electrons given $K$ incident photons follows a binomial distribution. In this specific case, the number of incident photons per laser pulse is sufficiently large for it to be approximated by a Poisson distribution.

One systematic effect in these measurements is the fluctuating laser intensity, which may vary by as much as $\pm\,20\%$ pulse to pulse, leading to similar fluctuations in the mean number of photoelectrons. Measurements were repeated at multiple pulse intensities, achieved with a tunable neutral density filter \cite{newsg_laser}. The data taken at each operating voltage was then divided into subsets based on the recorded PD amplitude of the events, within $\pm 5\%$ of a given value of PD amplitude. This ensured that the laser intensity in each subset was approximately constant, such that the deviation in the distribution of photoelectrons from a pure Poisson distribution is at most $0.1\%$ \cite{newsg_laser}. Other than this subdivision according to PD amplitude, no other selections were applied to the laser calibration data.

When $\mu \lesssim 1$, many events with no photoelectrons are expected, which are still recorded since the acquisition trigger is based on the PD channel. In this case, the observed signal is nominally a Gaussian distribution centered at $0\,\mathrm{ADU}$ due to the electronic noise in the $75\,\mathrm{\mu s}$ pulse integration window. However, laser events are sometimes---$\mathcal{O}(1\%)$---affected by pileup with the approximately $40$ to $55\,\mathrm{Hz}$ of radioactive source-induced events in the detector, in addition to the $10\,\mathrm{Hz}$ of laser pulses. Coincident pulses occurring randomly throughout the $2\,\mathrm{ms}$ event window can lead to events that are not reconstructed properly, in particular, if these pulses occur shortly before the integration window, affecting the computation of the reference baseline. This effect is evident from the asymmetric, non-Gaussian tails of the amplitude spectrum of 0 photoelectron events, as depicted in Fig.\ \ref{fig:wvalue_laser}. To properly account for this---which also impacts events with photoelectron pulses---the background spectrum, including pileup, was measured for every operating voltage. This was done by either collecting laser data with the fiber optic cable unplugged from the sphere to still trigger the DAQ via the PD, or by using data randomly triggered on the PD channel. Each measured spectrum was modeled with an adaptive bandwidth kernel density estimation (KDE) \cite{kde_article,kde_code,wang2011bandwidth,silverman} in order to generate a continuous background model, $P_{\mathrm{Noise}}$.

Combining all of the above, the laser spectrum model for a fixed laser intensity is given as

\begin{equation}
\begin{aligned}
    \mathcal{P}&\left(A | \mu, \theta, \left \langle G \right \rangle \right) = \left( \sum_{N=0}^{\infty} \Bigl[ P_{\mathrm{Poisson}}\left(N|\mu\right){\color{white}P_{\mathrm{I}}^{N}} \right. \\ & \left. {\color{white} \sum_0^0 } \times P_{\mathrm{Polya}}^{(N)} \left(A|\theta,\left \langle G \right \rangle \right) \Bigr] \right) \otimes P_{\mathrm{Noise}} \left(A \right),
\end{aligned}
\label{old_laser_model_eq}
\end{equation}

\noindent where $\otimes$ represents the convolution operator. In practice, the infinite sum is truncated at some $N$ when $P_{\mathrm{Poisson}}\left(N|\mu\right)$ becomes vanishingly small. For the case where $N=0$, $P_{\mathrm{Polya}}^{(0)}$ is treated as a Dirac delta function at $A$, such that the overall spectral contribution for $0$ electrons is simply $P_{\mathrm{Poisson}}\left(0|\mu\right) \times P_{\mathrm{Noise}} \left(A \right)$.


The laser calibration data was fit by maximizing the binned likelihood function for the data using the Markov chain Monte Carlo (MCMC) library \textsc{EMCEE} \cite{emcee}, yielding estimates of the model parameters. These fits were done with the ``stretch-move'' Metropolis-Hastings algorithm \cite{GW10,emcee}, propagating the random walkers for $10^4$ steps after an initial burn-in phase of $500$ steps, exceeding the recommendation that the chain be at least $50 \times$ as long as the autocorrelation time \cite{emcee}. 

An example of a single laser data spectrum and the resulting fit is shown in Fig.~\ref{fig:wvalue_laser}. The non-Gaussian shape of the null-event spectrum is apparent, as well as the near-exponential shape of the single electron component. This corresponds to a low value of $\theta$ for the Polya distribution ($0.64\,\pm\,0.06$ in this case). The measured mean gain was $103.6\,\pm\,1.5\,\mathrm{ADU/e^-}$ for this measurement, taken at an operating voltage of $1300\,\mathrm{V}$.

\begin{figure*}
\center
\includegraphics[width=0.9\textwidth]{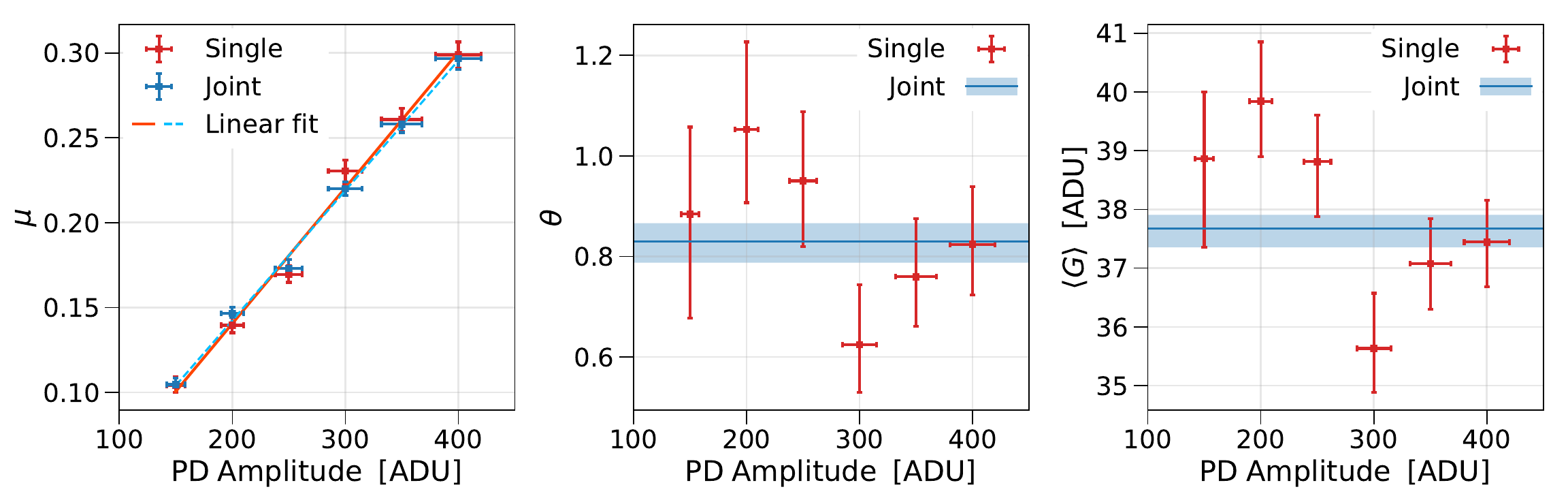}
  \caption{\label{fig:joint_laser} Laser calibration results for $\mu$, $\theta$, and $\left \langle G \right \rangle$ in subsets of PD amplitude, at an operating voltage of $1200 \,\mathrm{V}$. The fit results for each laser intensity subset are shown in red, and the joint fit in blue. Linear fits to the trends of $\mu$ vs.\ PD amplitude are also shown in the left panel. There is general agreement between the joint and individual fits.}
\end{figure*}

Individual fits were performed on each PD amplitude subset for all operating voltages as shown in Fig.~\ref{fig:joint_laser}. Additionally, a joint fit of all subsets (for a given voltage) was performed, using a common value of $\theta$ and $\left \langle G \right \rangle$, also shown in Fig.~\ref{fig:joint_laser}. This example, for data taken at $1200\,\mathrm{V}$, demonstrates that the distribution of photoelectrons varies linearly with laser intensity as expected. The joint-fit results are used in subsequent analyses. A summary of these results at each operating voltage is presented in Table \ref{tab:joint_laser}, with approximate $1\,\sigma$ statistical uncertainties defined to be the bounds of the MCMC samples where $\log \mathcal{L} \geq \mathrm{max} \{ \log \mathcal{L} \} - \frac{1}{2}$.

\begin{table}
\renewcommand*{\arraystretch}{1.7}
\begin{center}
    \begin{tabular}{ c | c | c }
    \hline \hline
    Operating voltage [$\mathrm{V}$] & $\left \langle G \right \rangle$ [$\mathrm{ADU/e^-}$] & $\theta$ \\ \hline
    $1200$ & $37.83_{-0.47}^{+0.07}$ & $0.84_{-0.07}^{+0.02}$ \\
    $1250$ & $60.44_{-0.23}^{+0.45}$ & $0.63_{-0.02}^{+0.03}$ \\
    $1300$ & $97.27_{-0.38}^{+0.42}$ & $0.60_{-0.02}^{+0.02}$ \\ \hline \hline
    \end{tabular}
    \protect\caption{ Laser calibration joint-fit results (with $1\,\sigma$ statistical uncertainties) combining data at different laser pulse intensities. As expected, the avalanche gain increases with anode voltage.}
    \label{tab:joint_laser}
\end{center}
\end{table}

\section{\label{sec:Wvalue}Electron capture decay modeling}
\label{sec:ar_decay}

\begin{figure}[t]
  \centering
  \includegraphics[width=0.48\textwidth]{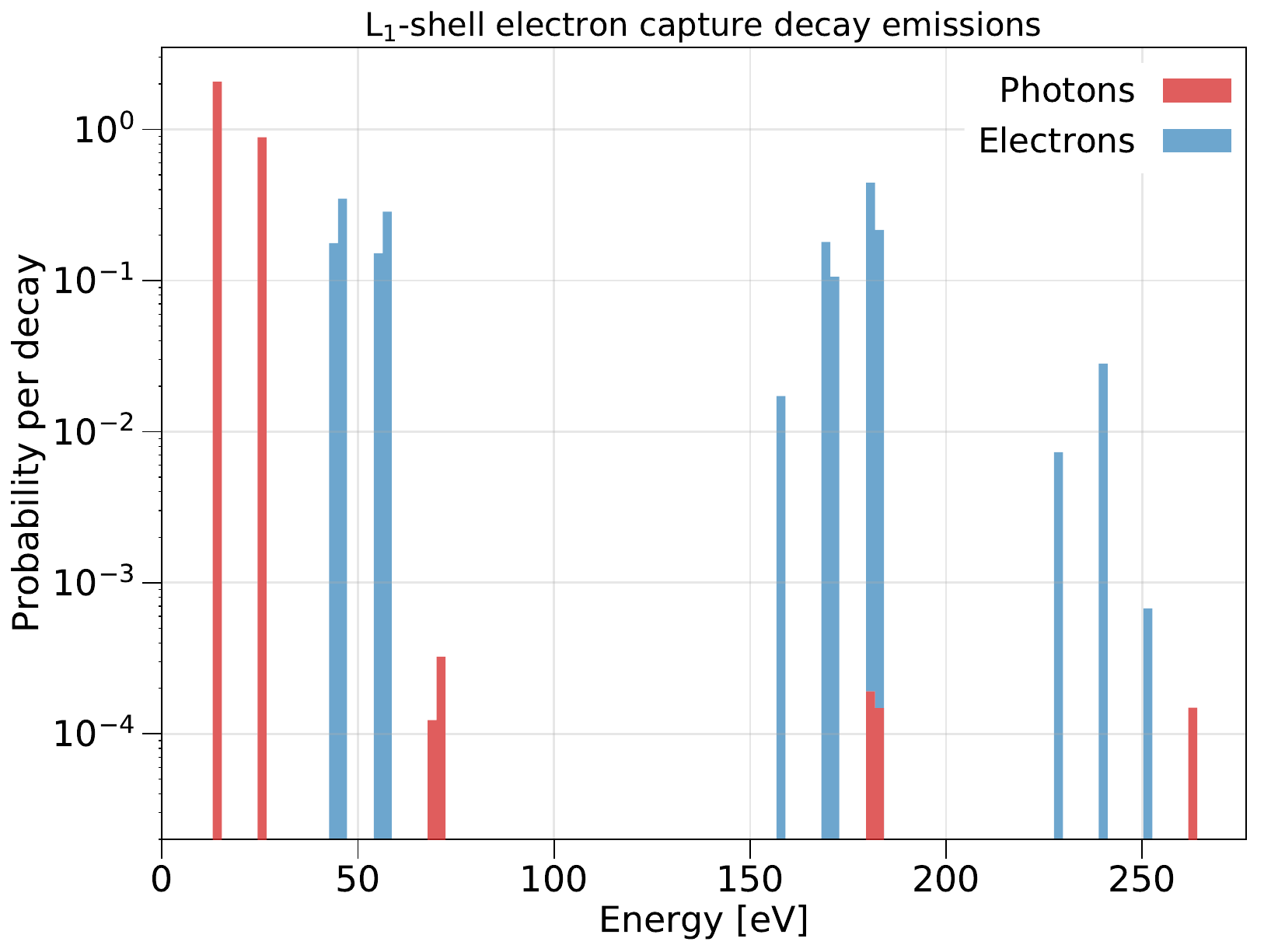}
  \caption{\label{fig:decay_spectrum} Spectrum of decay products (Auger electrons and x-rays) following the L$_1$-shell electron capture decay of $\mathrm{^{37}Ar}$ from $10^6$ simulated events.}
\end{figure}

Determining the mean ionization energy from the $\mathrm{^{37}Ar}$ calibration data requires a detailed model of its electron capture decay. The dominant decay channels for $\mathrm{^{37}Ar}$ are K-shell ($2.83\,\mathrm{keV}$) and L$_1$-shell ($277\,\mathrm{eV}$) electron capture, with branching ratios of $90.4\,\%$ and $8.4\,\%$, respectively, as given by the \textsc{BETASHAPE} code \cite{betashape}. Considering K-shell electron capture and the subsequent electron vacancy, there are many thousands of possible decay paths through which the daughter atom may reach its ground state. In the process, Auger electrons and x-rays of different energies are emitted; the spectrum of decay products following L$_1$-shell electron capture is shown in Fig. \ref{fig:decay_spectrum} as an example. While the total kinetic energy of the decay products is $2.83\,\mathrm{keV}$ (or $277\,\mathrm{eV}$ for L$_1$-shell capture), the mean number of ionization electrons produced in the gas may vary since $W$ is a function of energy. Further, different decay paths may produce varying numbers of initial electrons in the gas from Auger emissions, adding to the total number of electrons produced by ionization in the gas. Therefore, a simulation of the decay process is required.

Following the method of Ref.\ \cite{ringbom} and similar to Refs.\ \cite{Xe127_electronCapture,darkside_calib}, the simulation begins each iteration with a vacancy in either the K or L$_1$ shells. The vacancy is propagated as permissible emissions of Auger electrons or x-rays are selected at each step. The relative probabilities and energies of emitted particles in each cascade step are given by the RELAX code, using the EADL2017 library \cite{relax,eadl}. In many instances, additional vacancies are created as the excited state transitions to the ground state, e.g.\ through Auger electron emissions. Vacancies at each step are filled by order of their energy level, although the order of filling was found to have a negligible impact compared to filling them chronologically or randomly.

\begin{figure}
  \centering
  \includegraphics[width=0.48\textwidth]{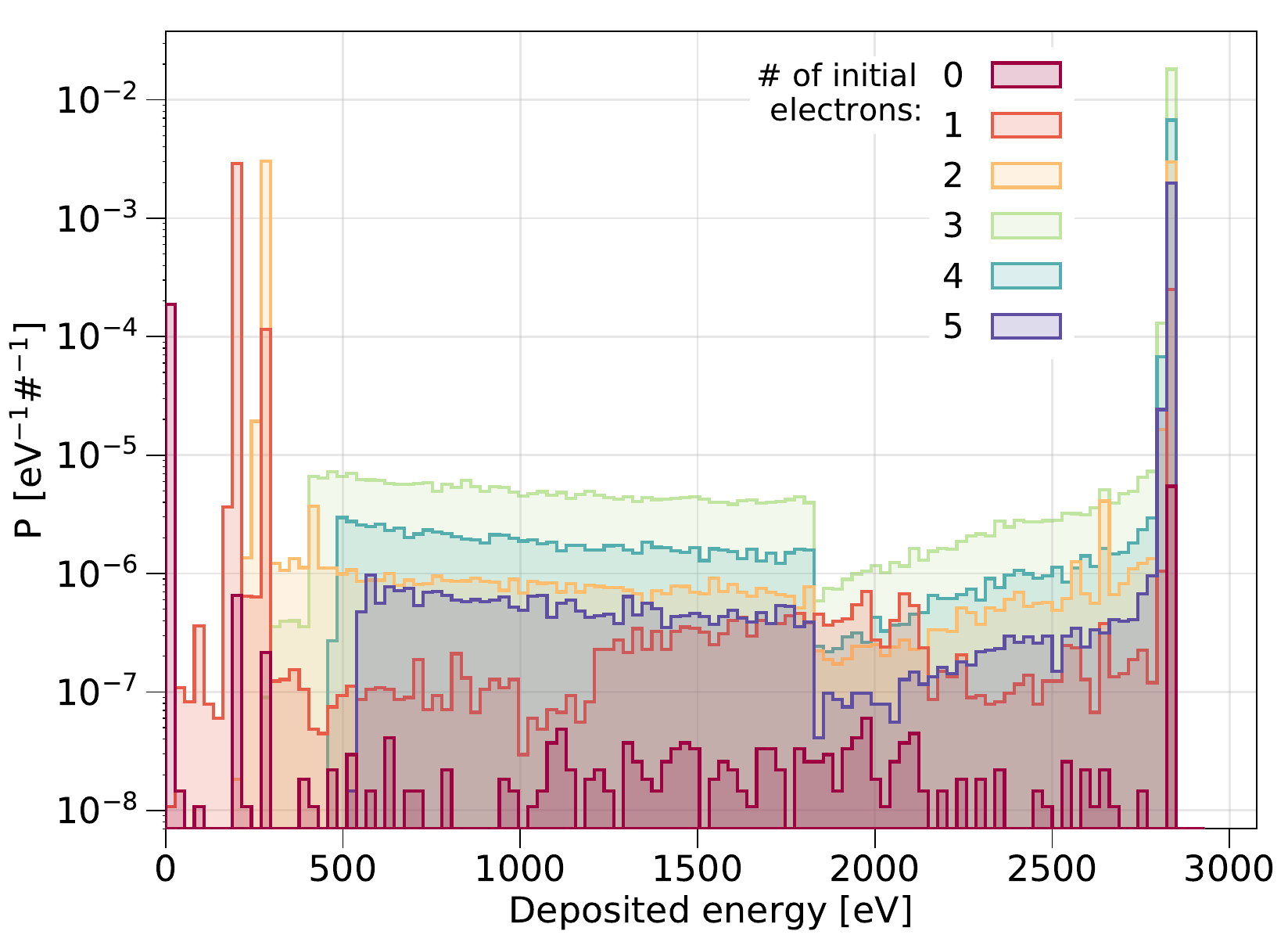}
  \caption{\label{fig:cascade} Histograms of the simulated energy deposited in the detector per $\mathrm{^{37}Ar}$ decay, for different numbers of initial (Auger) electrons, normalized with respect to energy and the number of initial electrons.}
\end{figure}

Due to the relatively small scale of the SPC and transparency of the gas, many of the $\mathcal{O}(1\, \mathrm{keV})$ energy photons may escape the detector without interaction, and electrons may only deposit some of their kinetic energy in the gas before reaching the detector wall. Therefore, \textsc{Geant4} simulations were used to calculate the spectrum of deposited energy within the gas volume by all possible decay cascade particles \cite{geant4} given the experimental setup. Specifically, the \textsc{Shielding} physics list was used \cite{shieldings}, with a production cutoff at $14\,\mathrm{eV}$. This yields a model for $\mathrm{^{37}Ar}$ decay in the SPC that is a function of the total energy deposited in the gas per event, as well as the number of initial Auger electrons in each decay cascade. This is depicted in Fig.~\ref{fig:cascade}. In addition to strong emissions at total energies of $277\,\mathrm{eV}$ and $2.83\,\mathrm{keV}$, there is a broad continuum in-between, with another substantial contribution at $200\,\mathrm{eV}$ from partially escaped K-shell decay cascades. A large number of $12.5\,\mathrm{eV}$ photons are produced by the capture of free electrons by the M shell as the final step to reaching the ground state, but these are not capable of ionization in this gas and are thus ignored.

\section{Ionization response measurements}
\label{sec:wvalue_fits}

The aluminum fluorescence and $\mathrm{^{37}Ar}$ calibration data were treated in the same way as the laser data described above. No pulse shape or other selections were applied to the datasets, except to exclude tracklike events with large risetimes (predominantly from cosmic-ray muons) and to remove UV laser events when it was operated concurrently. Analysis of the data was split into high and low-energy regimes in-between the dominant low-energy $\mathrm{^{37}Ar}$ peak and aluminum fluorescence peak due to the different treatment of systematics, described below. The UV laser calibration results given in Sec.~\ref{sec:laser} provided multivariate constraint terms (prior probabilities) for the mean gain $\left \langle G \right \rangle$, $\theta$ from the Polya distribution. 

The data was fit with a custom MCMC algorithm \cite{jin_thesis} and implemented through the library \textsc{EMCEE} \cite{emcee}. In this approach, the Metropolis--Hastings random walk is interrupted after approximately $10$ steps. The random walkers are then restarted along the previous maximum-likelihood boundary of the MCMC samples. This process is iterated until convergence is achieved, forcing more efficient mapping of the boundary of the likelihood function in high-dimensional parameter spaces \cite{pico_nr}. Approximate $1\,\sigma$ statistical error bars for visualization in plots and quoted numerical results are determined as described in Sec.~\ref{sec:laser}. In subsequent analyses, the MCMC samples obtained for each fit are used for parameter inference \cite{newsg_lsm}.

\subsection{Low energy analysis}
\label{subsec:low}

\begin{figure*}[tbh]
\center
\includegraphics[width=0.48\textwidth]{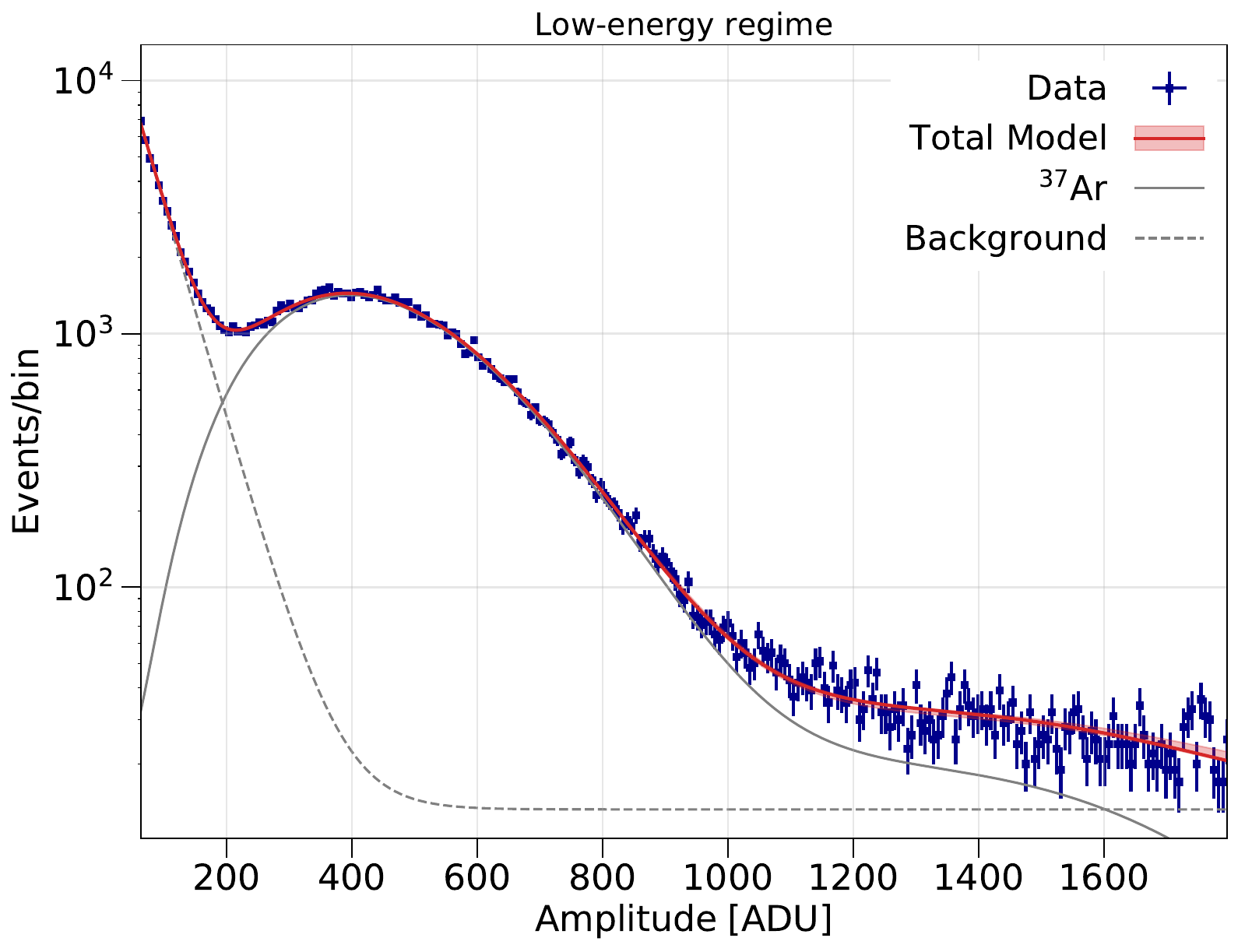}
\includegraphics[width=0.48\textwidth]{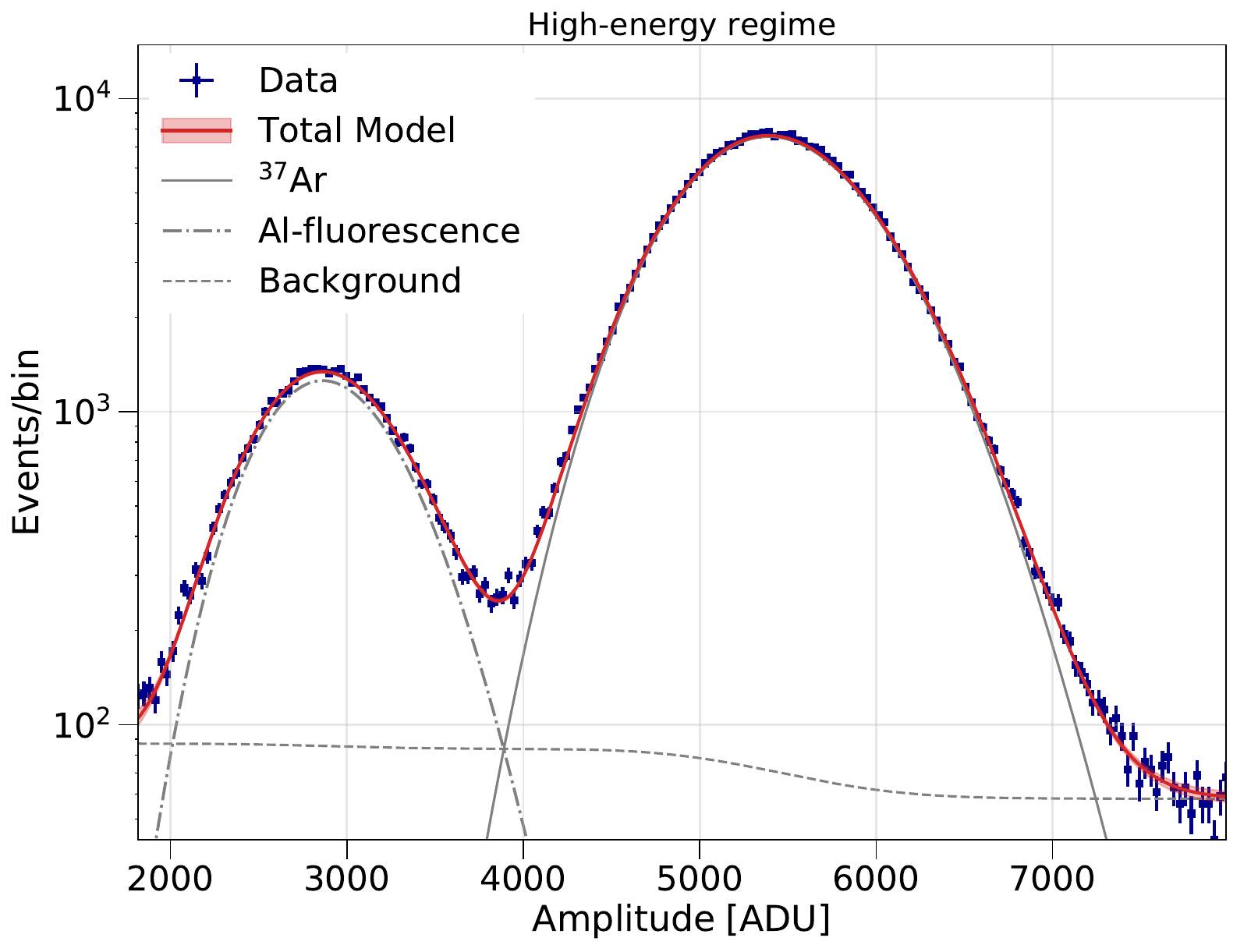}
  \caption{\label{fig:FitExamples} An example of low-energy $\mathrm{^{37}Ar}$ (left) and high-energy (right) calibration data taken at $1250\,\mathrm{V}$. The accompanying fits are shown in red, with $1\,\sigma$ uncertainty bands that are largely not visible. The components of each fit---$\mathrm{^{37}Ar}$, aluminum fluorescence, and backgrounds---are shown as gray curves with line styles indicated in the plot legends. The p-values for these fits are $0.14$ (low-energy) and $0.15$ (high-energy).}
\end{figure*}

The low-energy calibration data---an example of which is shown in Fig.\ \ref{fig:FitExamples}---consists of a continuum of energy depositions from $\mathrm{^{37}Ar}$ decay as discussed in Sec.~\ref{sec:ar_decay}, as well as a background of single or few-electron events. This data was fit in a ``forward-folding'' manner, beginning with the $\mathrm{^{37}Ar}$ cascade simulation to determine the probability of having $i$ electrons produced per decay---including $k$ initial Auger electrons. Then, the detector response to those $i$ electrons was applied, as determined in Sec.~\ref{sec:laser}.

The $\mathrm{^{37}Ar}$ simulation results (shown in Fig.\ \ref{fig:cascade}) are randomly sampled to obtain the deposited energy and number of initial electrons ($k$) per event. This contribution to the $i$ total electrons is fixed, while the $i-k$ electrons from ionizations in the gas fluctuate around a mean number $\mu$. From Eq.\ (\ref{eq:w_inokuti}), the mean number of ionizations induced by $M$ decay products each with energy $E_j$ is

\begin{equation}
    \label{eq:cascade_mean}
    \mu = \sum_j^M \frac{E_j}{W(E_j)} = \frac{1}{W_a} \left( \sum_j^M E_j - M \cdot U \right).
\end{equation}



\noindent Using this, a joint probability density function $P_k(\mu)$ was calculated for $\mu$ and $k$ initial electrons, for given values of $W_a$ and $U$.

The stochastic fluctuations of the $i-k$ ionization electrons around $\mu$ are modeled with the discrete Conway--Maxwell--Poisson (COM--Poisson) distribution \cite{mainCOM,newsg_fano}, the dispersion of which is described by the Fano factor $F$. In contrast to the Poisson-distributed ($F=1$) photoelectrons produced by the UV laser, ionization in the gas is more narrowly distributed with $F \approx 0.2$ \cite{uno2}, necessitating the use of a probability distribution with flexibility in $F$ \cite{newsg_fano}. 

Following the above, the probability distribution for having $i$ total electrons is given as 

\begin{equation}
\label{eq:low_pi}
   P_{\mathrm{Low}}(i|\mu, F) = \sum_{k=0}^{k=5} \int P_{\mathrm{COM}}(i-k|\mu,F) P_k(\mu) d \mu.
\end{equation}

\noindent This is then multiplied by the $i\mathrm{th}$ convolution of the Polya distribution to give the detector response to $i$ electrons, which are then convolved with the baseline noise spectrum. 

A background contribution contribution was also included in the fitting model, consisting of a uniform and an exponential component. The large, low-energy background described by the latter is likely composed of single electrons, poorly reconstructed events, and nonphysical pulses. However, no pulse-shape discrimination selections are applied to the data to avoid distorting the $\mathrm{^{37}Ar}$ spectrum. A significant fraction of this low-amplitude signal is induced by the relatively high rate of calibration source events themselves, precluding a separate background-only measurement. Using this background model and Eq.\ (\ref{eq:low_pi}), the complete model for the energy response of low-energy $\mathrm{^{37}Ar}$ events can be written as

\begin{equation}
\begin{aligned}
    \mathcal{P}&\left(A \right) = \left[ \sum_i  P(i|\mu, F)_{\mathrm{Low}} \times P_{\mathrm{Polya}}^{\mathrm{(i)}}(A|\left \langle G \right \rangle, \theta) \right] \\ &{\color{white} \sum_0^0 } \otimes P_{\mathrm{Noise}} \left(A \right) + B_{\mathrm{Flat}} + B_1 e^{-B_2 A}.
\end{aligned}
\label{eq:low_model}
\end{equation}


In addition to constraints on each fit from the corresponding UV laser measurements given in Sec.~\ref{sec:laser}, the high-energy calibration data was fit first (as described in Sec.~\ref{subsec:high}), the results of which provide a uniform prior probability distribution for $W_a$ for the low-energy fit. This corresponds to the maximum systematic range calculated for this parameter. The fit of the low-energy calibration data taken at $1250\,\mathrm{V}$ is shown in Fig.~\ref{fig:FitExamples} (left) as an example.

\subsection{High energy analysis}
\label{subsec:high}

The aluminum fluorescence peak and high energy $\mathrm{^{37}Ar}$ contributions were fit together, as their spectra overlap significantly. A similar model as the one described in Sec.~\ref{subsec:low} was used, with the $\mathrm{^{37}Ar}$ decay simulation again providing a joint probability distribution of the mean number of ionization electrons ($\mu_{\mathrm{Ar}}$) and $k$ initial electrons, $P_k(\mu_{\mathrm{Ar}})$, for that source. In this case, given the high energy of the decay products, a single parameter $W$ was used, neglecting the energy dependence. The aluminum fluorescence decay was considered to be monoenergetic and thus was described by a single value of $\mu_{\mathrm{Al}}$, with a rate ratio $R$ compared to the $\mathrm{^{37}Ar}$ peak. The probability distribution of $i$ total electrons in the high-energy regime is therefore

\begin{equation}
\label{eq:high_pi}
    \begin{aligned}
    P_{\mathrm{High}} & \left( i|\mu_{\mathrm{Ar}}, F_{\mathrm{Ar}}, \mu_{\mathrm{Al}}, F_{\mathrm{Al}} \right) = (1-R)\times \\ & \left( \sum_{k=0}^{k=5} \int P_{\mathrm{COM}}(i-k|\mu_{\mathrm{Ar}},F_{\mathrm{Ar}}) P_k(\mu_{\mathrm{Ar}}) d \mu_{\mathrm{Ar}} \right) \\ &+ R \times P_{\mathrm{COM}}\left(i|\mu_{\mathrm{Al}},F_{\mathrm{Al}} \right).
    \end{aligned}
\end{equation}

Although the $\mathrm{^{37}Ar}$ electron capture decay model discussed in Sec.~\ref{sec:ar_decay} predicts a continuum of energy depositions near the main peak at $2.83\,\mathrm{keV}$, in this region it is completely dominated by the aluminum fluorescence data. Therefore, a uniform-energy background component $B(A)$ was added to the fit, with independent background levels below and above the $^{37}\mathrm{Ar}$ peak ($B_1$ and $B_2$), joined smoothly by the reciprocal cumulative distribution function of the source model $\mathcal{P}(A)$, which can be written as 

\begin{equation}
    \label{eq:high_bkgd}
    B(A) = B_1 \times \left[1-\int_0^A \mathcal{P}(A')dA' \right] + B_2.
\end{equation}

This was done to account for the larger background component found below the fluorescence peak, likely a combination of Compton scattering and poorly reconstructed events. Altogether, the fit model for the high-energy calibration data (without the background component) can be written in terms of the contribution from $i$ total electrons from both the $\mathrm{^{37}Ar}$ and aluminum, each with their respective single value or spectrum of mean ionization electrons $\mu$ and separate Fano factors

\begin{equation}
\begin{aligned}
    \mathcal{P}\left(A \right) &= \left[ \sum_i P_{\mathrm{High}} \left( i|\mu_{\mathrm{Ar}}, F_{\mathrm{Ar}}, \mu_{\mathrm{Al}}, F_{\mathrm{Al}} \right) \right. \\ & \left. {\color{white} \sum_0^0 } \times P_{\mathrm{Polya}}^{\mathrm{(i)}}(A|\left \langle G \right \rangle, \theta ) \right]\otimes P_{\mathrm{Noise}} \left(A \right).
\end{aligned}
\label{eq:high_model}
\end{equation}

\begin{figure*}[tbh]
\center
\includegraphics[width=0.58\textwidth]{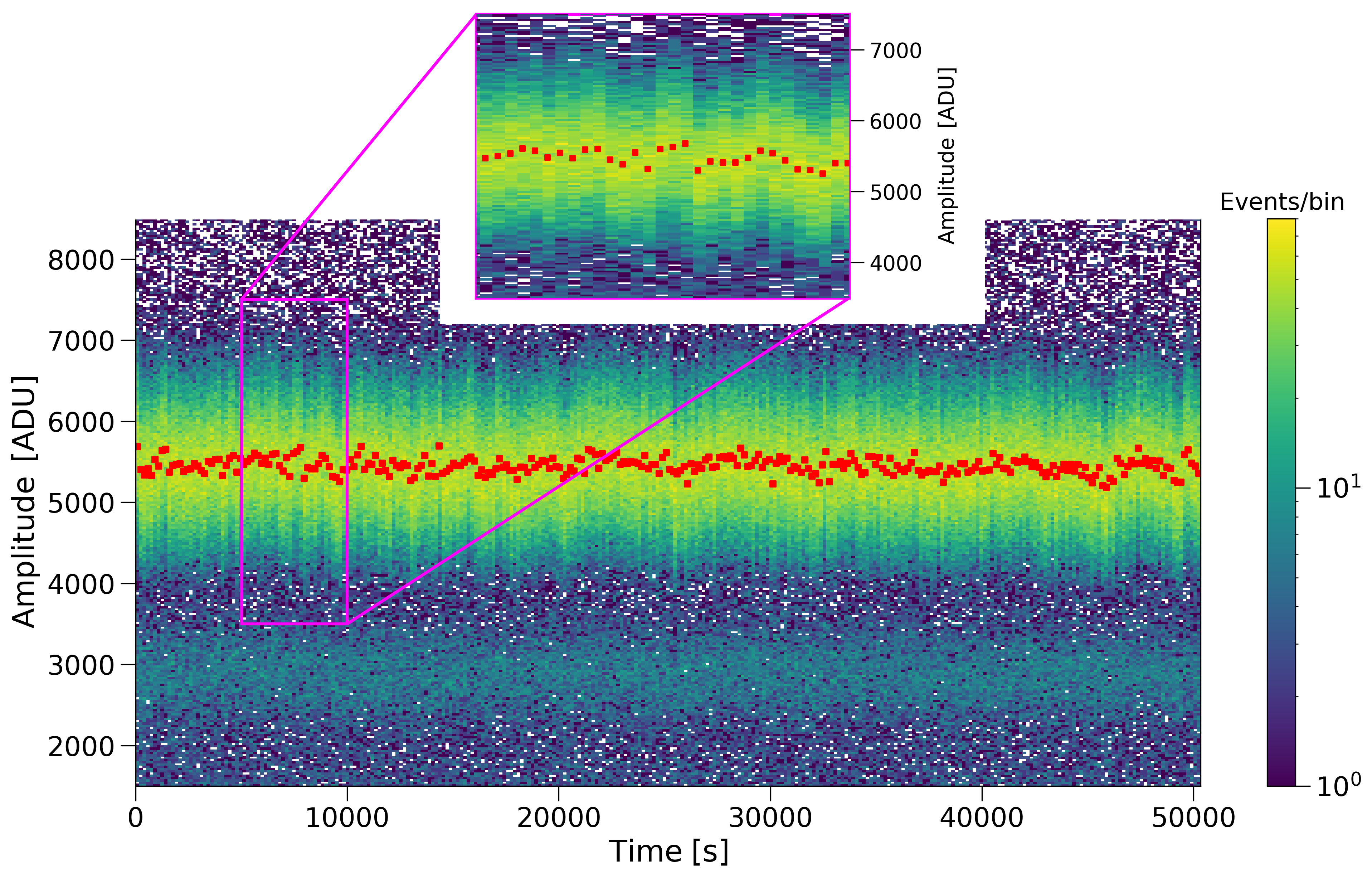}
\includegraphics[width=0.38\textwidth]{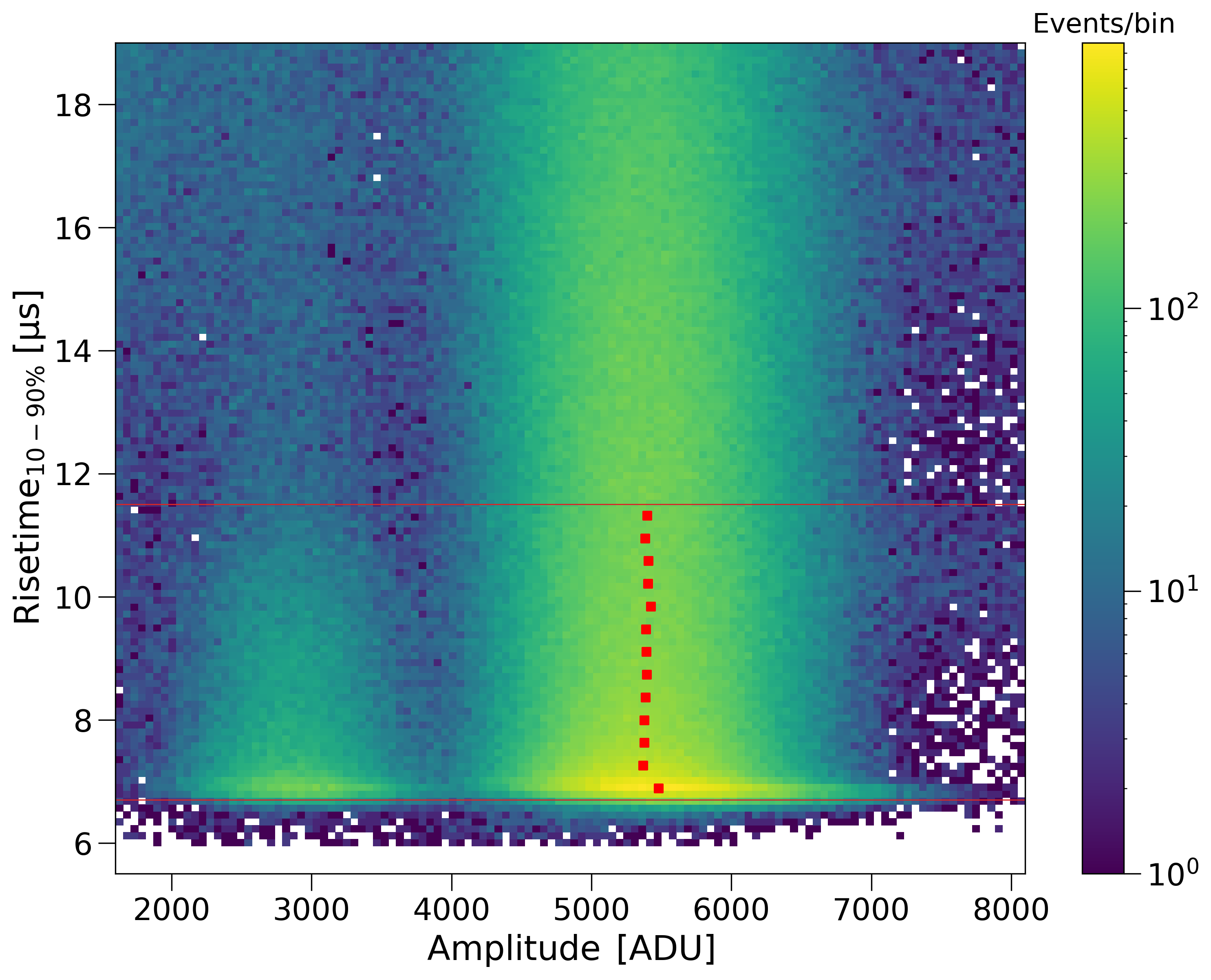}
  \caption{\label{fig:systematics} Left: amplitude vs.\ time of high-energy calibration data taken at $1250\,\mathrm{V}$, with the mean amplitude in individual time slices labeled with red points. The inset shows fluctuations in amplitude on timescales of $\mathcal{O}(1000\mathrm{s})$. Right: risetime vs.\ amplitude of the same calibration data showing the mean amplitude in slices of risetime (red points) inside the analysis range (red lines).}
\end{figure*}

As mentioned in Sec.~\ref{sec:setup}, one potential issue with the data is detector instabilities when operating at relatively high voltages. This may result in fluctuations of the detector gain and, therefore, of the amplitude of physics events beyond expected statistical fluctuations. These fluctuations distort event amplitudes on the order of a few percent over timescales of thousands of seconds, as seen in Fig.~\ref{fig:systematics} (left). A $\mbox{Lomb-Scargle}$ periodogram \cite{scargle,gregory} of the amplitude data over time revealed that the power of oscillations was marginally highest for periods of approximately $500$ to $5000\,\mathrm{s}$, motivating the time binning used in  Fig.~\ref{fig:systematics} (left). To address this as a systematic, the mean amplitude of high-energy $\mathrm{^{37}Ar}$ events was calculated in bins of 167 seconds. To avoid this broadening in the amplitude distribution impacting estimation of the Fano factor, the amplitude in each bin was translated to match the mean amplitude in the first bin for fitting. The maximum relative difference between the bin averages was propagated to give a range of corresponding values of $W_a$. This full systematic range was conservatively included in the final reported result, to remain agnostic about the direction and nature of the distortions.

Another potential systematic effect with a similar result may be electron attachment, the stochastic process by which some electrons are captured by electro-negative gas constituents (primarily water and oxygen) as they drift toward the anode \cite{Arnaud2018}. This effect would be stronger for events originating farther from the anode, leading to an anticorrelation between risetime and amplitude. In all datasets used, there was no evidence of electron attachment at high risetimes corresponding to surface events. However, a slight increase in the amplitude of the lowest risetime events was observed for both high energy peaks in all datasets, which may be an effect of the pulse treatment or some other unaccounted for effect. 

The corresponding systematic uncertainty was estimated using a similar procedure by taking the average amplitude of the high-energy $\mathrm{^{37}Ar}$ peak in bins of risetime, from the minimum risetime in each data set (typically around 4--5 $\mathrm{\mu s}$) to 12 $\mathrm{\mu s}$, as shown in Fig.~\ref{fig:systematics} (right). As before, the conservative choice was made to propagate the maximum extent of this systematic effect to a corresponding range of values of $W$. The fit of the high-energy calibration data taken at $1250\,\mathrm{V}$---``corrected'' for the systematic effects described above---is shown in Fig.~\ref{fig:FitExamples} (right) as an example. 

\subsection{Fit results}
\label{ss:results}

\begin{table}
\renewcommand*{\arraystretch}{1.7}
\begin{center}
    \begin{tabular}{ c | c | c | c }
    \hline \hline
    Anode HV [V] & $W_a$ [eV] & $U$ [eV] & $F$ \\ \hline
    $1200$ & $32.1_{-1.7}^{+0.7}$ & $11.7_{-1.5}^{+2.7}$ & $0.51_{-0.04}^{+0.03}$ \\
    $1250$ & $31.1_{-0.5}^{+0.9}$ & $14.6_{-1.4}^{+0.9}$ & $0.43_{-0.02}^{+0.01}$ \\
    $1300$ & $33.1_{-1.5}^{+0.1}$ & $11.7_{-0.4}^{+2.4}$ & $0.61_{-0.03}^{+0.01}$ \\ \hline \hline
    \end{tabular}
    \protect\caption{Low energy analysis results at all anode voltages, with independent $1\,\sigma$ statistical uncertainties for each parameter.}
    \label{tab:low}
\end{center}
\end{table}

\begin{figure}
  \centering
  \includegraphics[width=0.48\textwidth]{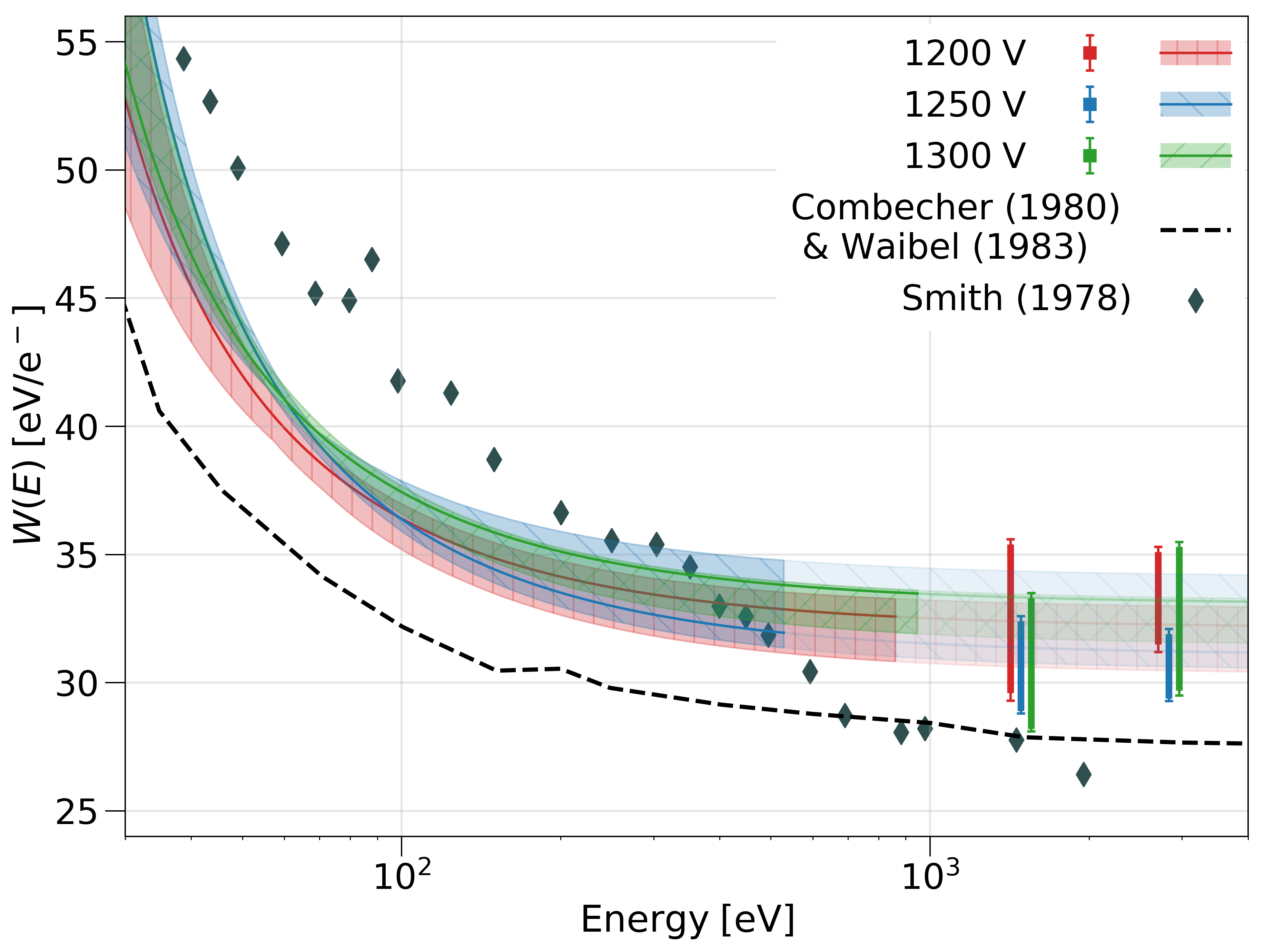}
  \caption{\label{fig:combinedResult} Combined results for $W(E)$ from calibration data taken at three different detector gains (voltages) in $50\,\mathrm{mbar}$ methane. The data points with error bars represent the results from aluminum fluorescence and high-energy $\mathrm{^{37}Ar}$ with systematic (thick markers) and statistical uncertainties (ticked error bars) from the fit of the systematic-corrected data. The curves and shaded bands represent the low-energy calibration best fits and joint $1\,\sigma$ uncertainty from $W_a$ and $U$, which terminate at the end of their respective fitting ranges with a faint extrapolation to compare with the high-energy results. The combined historical results of Combecher (1980) \cite{combecher} and Waibel and Grosswendt (1983) \cite{waibel1983} are shown (dashed black line), as well as the result of Smith and Booz (1978) \cite{booz} (diamond markers).}
\end{figure}

The key fit results from the low-energy regime ($W_a$, $U$, and $F$) are presented in Table \ref{tab:low}, and those of the high-energy regime ($W_a$ and $F$ for both peaks) are given in Table \ref{tab:high}. As described in Sec.~\ref{subsec:high}, the high-energy results for $W_a$ are reported conservatively including the maximum extent of the systematic effects considered, with that full range of the parameter value considered to be equally probable. Statistical uncertainties for $W_a$ in the high-energy regime were on the order of $0.15\,\mathrm{eV}$, but are not listed as they only apply to the fit of the systematic-corrected data, and are small relative to the systematic uncertainty. All other parameter results are listed with $1\,\sigma$ statistical uncertainties, approximately determined to be the extent of each parameter where $\log \mathcal{L} \geq \mathrm{max}(\log \mathcal{L}) - \frac{1}{2}$. This also incorporates uncertainties from the UV laser analysis results on $\theta$ and $\left \langle G \right \rangle$. The combined ionization yield measurements from the high and low-energy analyses are presented in Fig.~\ref{fig:combinedResult}.

\begin{table}
\renewcommand*{\arraystretch}{1.7}
\begin{center}
    \begin{tabular}{ c | c | c | c }
    \hline \hline
    Operating voltage [V] & Peak & $W_a$ [eV] & $F$ \\ \hline
    $1200$ & $\mathrm{Alum.}$ & $\left(29.6 - 35.4 \right)$ & $0.33_{-0.01}^{+0.01}$ \\
    $1200$ & $\mathrm{^{37}Ar}$ & $\left(31.5 - 35.1 \right)$ & $0.39_{-0.01}^{+0.01}$ \\
    & & & \\
    $1250$ & $\mathrm{Alum.}$ & $\left(28.9 - 32.4 \right)$ & $0.33_{-0.02}^{+0.02}$ \\
    $1250$ & $\mathrm{^{37}Ar}$ & $\left(29.4 - 31.9 \right)$ & $0.30_{-0.01}^{+0.01}$ \\
    & & & \\
    $1300$ & $\mathrm{Alum.}$ & $\left(28.2 - 33.3 \right)$ & $0.49_{-0.02}^{+0.02}$ \\
    $1300$ & $\mathrm{^{37}Ar}$ & $\left(29.7 - 35.3 \right)$ & $0.43_{-0.01}^{+0.01}$ \\ \hline \hline
    \end{tabular}
    \protect\caption{Fit results of the high-energy peaks at all anode voltages, including the systematic range on the value of $W_a$, and $1\,\sigma$ statistical uncertainties on $F$.}
    \label{tab:high}
\end{center}
\end{table}

The ionization yield results for each measurement are roughly inline with expectations, in terms of both the high-energy limiting value $W_a$ and behavior at low energy (see Sec.~\ref{sec:discussion} for further discussion). As expected, no evidence of a dependence with operating voltage was observed (see Fig.~\ref{fig:combinedResult}). The results for the Fano factor are generally higher than expected from the literature \cite{grosswendt_1985}, with values ranging from roughly $0.3$ to $0.6$ (see Tables \ref{tab:low} and \ref{tab:high}) compared to the expected value of approximately $0.2$ at an energy of $1\,\mathrm{keV}$. This is likely due to resolution-broadening effects specific to SPCs that are not otherwise accounted for in the fit model. Consequently, these Fano factor measurements---and the similar $\mathrm{^{37}Ar}$ calibration performed at the LSM \cite{newsg_lsm}---are treated as upper limits on the intrinsic Fano factor of methane, while still accurately characterizing the detector's energy resolution. The behavior of $F$ with energy presented here is generally consistent with expectations from the literature, with a similar evolution as $W(E)$ \cite{uno2, grosswendt_1985}.

\section{Discussion and conclusions}
\label{sec:discussion}

Measurements of the ionization yield of methane gas for electronic recoils are presented. These were performed with a $30\,\mathrm{cm}$ diameter SPC, benefiting from \emph{in situ} UV laser calibrations. Data was collected at three different operating voltages. The calibration sources used---$\mathrm{^{37}Ar}$ and aluminum fluorescence---yielded measurements of this quantity across an energy range of approximately $50\,\mathrm{eV}$ to $2.8\,\mathrm{keV}$, the regime-of-interest for low mass DM searches performed by the $\mbox{NEWS-G}$ collaboration. These measurements provide independent validation of a similar calibration performed with the $\mbox{NEWS-G}$ detector installed at the LSM, also operated with methane gas \cite{newsg_lsm}.

The measured quantity $W_a$ may be compared with the high-energy limiting value of the ``W-value'' \cite{inokuti}. When compared to earlier measurements of the W-value in methane, the results obtained here are typically about $15\%$---or $4.5\,\mathrm{eV}$---higher than the combined measurements of Combecher (1980) \cite{combecher} and Waibel and Grosswendt (1983) \cite{waibel1983} throughout the relevant energy range. In comparison, the average systematic uncertainty range for $W_a$ in this work is approximately $4\,\mathrm{eV}$ in the high-energy regime. Regardless, given the \emph{in situ} energy response characterization provided by these measurements, the results presented here are appropriate to describe the energy response of SPCs with methane. Thus, this measurement was used as a calibration prior for a recent NEWS-G dark matter analysis \cite{newsg_lsm}.

\begin{acknowledgments}
We thank Davide Franco (Laboratoire Astroparticule et Cosmologie) for helpful discussions regarding the electron capture decay simulations conducted by the DarkSide collaboration. 
This research was undertaken, in part, thanks to funding from the Canada Excellence Research Chairs Program, the Canada Foundation for Innovation, the Arthur B. McDonald Canadian Astroparticle Physics Research Institute, Canada, the French National Research Agency (ANR-15-CE31-0008), and the Natural Sciences and Engineering Research Council of Canada. This project has received support from the European Union's Horizon 2020 research and innovation programme under grant agreements No.~841261 (DarkSphere), No.~845168 (neutronSPHERE), and No.~101026519 (GaGARin).
Support from the U.K. Research and
Innovation---Science and Technology Facilities Council (UKRI-STFC), through grants No.~ST/V006339/1, No.~ST/S000860/1, No. ST/W000652/1, No. ST/X005976/1, and No. ST/X508913/1, the UKRI Horizon Europe Underwriting scheme (GA101066657/Je-S EP/X022773/1), and the Royal Society International Exchanges Scheme (IES$\backslash$R3$\backslash$170121) is acknowledged.
Support by the Deutsche Forschungsgemeinschaft (DFG, German Research Foundation) under Germany’s Excellence Strategy---EXC 2121 ``Quantum Universe''---390833306 is acknowledged.
The work of D.~Durnford was additionally supported by the NSERC Canada Graduate Scholarships---Doctoral program (CGSD).
\end{acknowledgments}


\bibliography{Wvalue}

\end{document}